\documentstyle[twoside]{article}

\catcode`\@=11
\long\def\@makefntext#1{
\protect\noindent \hbox to 3.2pt {\hskip-.9pt  
$^{{\eightrm\@thefnmark}}$\hfil}#1\hfill}		%CAN BE USED 

\def\@makefnmark{\hbox to 0pt{$^{\@thefnmark}$\hss}}	%ORIGINAL 
	
\def\ps@myheadings{\let\@mkboth\@gobbletwo
\def\@oddhead{\hbox{}
\rightmark\hfil\eightrm\thepage}   
\def\@oddfoot{}\def\@evenhead{\eightrm\thepage\hfil
\leftmark\hbox{}}\def\@evenfoot{}
\def\sectionmark##1{}\def\subsectionmark##1{}}

\oddsidemargin=\evensidemargin
\newcounter{sectionc}\newcounter{subsectionc}\newcounter{subsubsectionc}
\renewcommand{\section}[1] {\vspace{12pt}\addtocounter{sectionc}{1} 
\setcounter{subsectionc}{0}\setcounter{subsubsectionc}{0}\noindent 
	{\tenbf\thesectionc. #1}\par\vspace{5pt}}
\renewcommand{\subsection}[1] {\vspace{12pt}\addtocounter{subsectionc}{1} 
	\setcounter{subsubsectionc}{0}\noindent 
	{\bf\thesectionc.\thesubsectionc. {\kern1pt \bfit #1}}\par\vspace{5pt}}
\renewcommand{\subsubsection}[1] {\vspace{12pt}\addtocounter{subsubsectionc}{1}
	\noindent{\tenrm\thesectionc.\thesubsectionc.\thesubsubsectionc.
	{\kern1pt \tenit #1}}\par\vspace{5pt}}
\newcommand{\nonumsection}[1] {\vspace{12pt}\noindent{\tenbf #1}
	\par\vspace{5pt}}

\newcommand{\textlineskip}{\baselineskip=13pt}
\newcommand{\smalllineskip}{\baselineskip=10pt}

\def\eightcirc{
\begin{picture}(0,0)
\put(4.4,1.8){\circle{6.5}}
\end{picture}}
\def\eightcopyright{\eightcirc\kern2.7pt\hbox{\eightrm c}} 

\newcommand{\copyrightheading}[1]
	{\vspace*{-2.5cm}\smalllineskip{\flushleft
        {\footnotesize Los Alamos archive: gr-qc/9412033r #1}\\
        {\footnotesize $\eightcopyright$\, H.C. Rosu (1997)}\\
	 }}

\def\abstracts#1#2#3{{
	\centering{\begin{minipage}{4.5in}\baselineskip=10pt\footnotesize
	\parindent=0pt #1\par 
	\parindent=15pt #2\par
	\parindent=15pt #3
	\end{minipage}}\par}} 

\renewenvironment{thebibliography}[1]
	{\frenchspacing
	 \ninerm\baselineskip=11pt
	 \begin{list}{\arabic{enumi}.}
        {\usecounter{enumi}\setlength{\parsep}{0pt}     
	 \setlength{\leftmargin 12.7pt}{\rightmargin 0pt} %FOR 1--9 ITEMS
         \setlength{\itemsep}{0pt} \settowidth
	{\labelwidth}{#1.}\sloppy}}{\end{list}}

\newcounter{itemlistc}
\newcounter{romanlistc}
\newcounter{alphlistc}
\newcounter{arabiclistc}

\def\@citex[#1]#2{\if@filesw\immediate\write\@auxout
	{\string\citation{#2}}\fi
\def\@citea{}\@cite{\@for\@citeb:=#2\do
	{\@citea\def\@citea{,}\@ifundefined
	{b@\@citeb}{{\bf ?}\@warning
	{Citation `\@citeb' on page \thepage \space undefined}}
	{\csname b@\@citeb\endcsname}}}{#1}}

\newif\if@cghi
\def\cite{\@cghitrue\@ifnextchar [{\@tempswatrue
	\@citex}{\@tempswafalse\@citex[]}}
\def\citelow{\@cghifalse\@ifnextchar [{\@tempswatrue
	\@citex}{\@tempswafalse\@citex[]}}
\def\@cite#1#2{{$\null^{#1}$\if@tempswa\typeout
	{IJCGA warning: optional citation argument 
	ignored: `#2'} \fi}}

\def\@refcitex[#1]#2{\if@filesw\immediate\write\@auxout
	{\string\citation{#2}}\fi
\def\@citea{}\@refcite{\@for\@citeb:=#2\do
	{\@citea\def\@citea{, }\@ifundefined
	{b@\@citeb}{{\bf ?}\@warning
	{Citation `\@citeb' on page \thepage \space undefined}}
	\hbox{\csname b@\@citeb\endcsname}}}{#1}}

\def\@refcite#1#2{{#1\if@tempswa\typeout
        {IJCGA warning: optional citation argument
	ignored: `#2'} \fi}}

\def\refcite{\@ifnextchar[{\@tempswatrue
	\@refcitex}{\@tempswafalse\@refcitex[]}}

\def\pmb#1{\setbox0=\hbox{#1}
	\kern-.025em\copy0\kern-\wd0
	\kern.05em\copy0\kern-\wd0
	\kern-.025em\raise.0433em\box0}

\def\fnt#1#2{\footnotetext{\kern-.3em
	{$^{\mbox{\scriptsize #1}}$}{#2}}}

\font\tenrm=cmr10
\font\tenit=cmti10 
\font\tenbf=cmbx10
\font\bfit=cmbxti10 at 10pt
\font\ninerm=cmr9

\font\eightrm=cmr8

\def\qed{\hbox{${\vcenter{\vbox{			%HOLLOW SQUARE
   \hrule height 0.4pt\hbox{\vrule width 0.4pt height 6pt
   \kern5pt\vrule width 0.4pt}\hrule height 0.4pt}}}$}}

\begin{document}

%\runninghead{Instructions for Typesetting Camera-Ready
%Manuscripts $\ldots$} {Instructions for Typesetting Camera-Ready
%Manuscripts $\ldots$}

%Comment (H. Rosu): produce fraza de mai sus la inceputul fiecarei pagini

\normalsize\textlineskip
\thispagestyle{empty}
\setcounter{page}{1}

\copyrightheading{}			%{Vol. 0, No.0 (1992) 000--000}

\vspace*{0.88truein}

%\fpage{1} %%%%%%%%%%%%%%%%%%%%%%%%%%%%%%%%%%%%%%%%%%%%%%%%%%%%%%%%%%%
\centerline{\bf NOTES ON SYNCHROTRON RADIATION}
\vspace*{0.035truein}
%\centerline{\bf MANUSCRIPTS USING COMPUTER SOFTWARE\footnote{For
%the title, try not to use more than 3 lines. Typeset the title
%in 10 pt Times Roman, uppercase and boldface.}}
\vspace*{0.37truein}
\centerline{\footnotesize HARET C. ROSU}
%\footnote{Typeset names in
%10 pt Times Roman, uppercase. Use the footnote to indicate the
%present or permanent address of the author.}}
\vspace*{0.015truein}
\centerline{\footnotesize\it Instituto de F\'{\i}sica,
Universidad de Guanajuato, Apdo Postal E-143, Le\'on, Gto, Mexico}
\baselineskip=10pt
%\centerline{\footnotesize\it City, State ZIP/Zone,
%Country\footnote{State completely without abbreviations, the
%affiliation and mailing address, including country. Typeset in 8
%pt Times Italic.}}
\vspace*{10pt}
%\centerline{\footnotesize SECOND AUTHOR}
%\vspace*{0.015truein}
%\centerline{\footnotesize\it Group, Laboratory, Address}
%\baselineskip=10pt
%\centerline{\footnotesize\it City, State ZIP/Zone, Country}
\vspace*{0.225truein}
%\publisher{(May 27, 1997)}{(December 29, 1997)}

\vspace*{0.21truein}
\abstracts{I comment on a number of theoretical issues
related to magnetobremsstrahlung, and especially on synchrotron radiation
and Unruh (temperature) radiation, that I consider of importance for the
current progress towards a better understanding of the stationary features
of such fundamental
radiation patterns both in an accelerator context and, more generally,
in the physical world.}{}{}

%\vspace*{10pt}
%\keywords{The contents of the keywords}

\textlineskip                  %) USE THIS MEASUREMENT WHEN THERE IS
\vspace*{12pt}                 %) NO SECTION HEADING

\vspace*{1pt}\textlineskip	%) USE THIS MEASUREMENT WHEN THERE IS
%\section{General Appearance}    %) A SECTION HEADING
\vspace*{-0.5pt}
\noindent

%%%%%%%%%%%%%%%%%%%%%%%%%%%%%%%%%%%%%%%%%%%%%%
%PACS number(s):  98.80.Hw, 11.30.Pb
%\vskip 2cm

\noindent
%%%%%%%%%%%%%%%%%%%%%%%%%%%%%%%%%%%%%%%%%%%%%%%%%%%%%%%%%%%%%%%%%%%%%

%\newpage

%\pagebreak

%\textheight=7.8truein
%\setcounter{footnote}{0}
%\renewcommand{\thefootnote}{\alph{footnote}}

%\section{The Main Text}
\noindent

%\bigskip
%PACS numbers:  41.60.Ap, 41.75.Ht, 05.40.+j

%\vskip 2cm

%\vskip 1cm

%IFUG-28/94 [$\cal H \cal C \cal R$]
%Initial title:
%Preliminary notes on synchrotron radiation as temperature radiation

%\vskip 1cm

%gr-qc/9412033 of December 12, r-Dec. 13, 1994; December 1997

%\vskip 1cm

%e-mail: rosu@ifug.ugto.mx

% Draft as of December 28, 1997

1. Introduction

2. Magnetobremsstrahlung radiation patterns

3. Quantum criteria in synchrotron radiation

4. Trajectories (worldlines) and radiation patterns from quantum detectors

5. QED bremsstrahlung and vacuum thermal noises

6. Radiometric characterization of synchrotron radiation

7. The problem of coherence in synchrotron radiation

8. Bremsstrahlung and continued fractions

9. Conclusions

\newpage
% 1.  %%%%%%%%%%%%%%%%%%%%%%%%%%%%%%%%%%%%%%%%%%%%%%%%%%%%%%%%%%%%%%%%%%%%
\section{Introduction}
%%%%%%%%%%%%%%%%%%%%%%%%%%%%%%%%%%%%%%%%%%%%%%%%%%%%%%%%%%%%%%%%%%%%%%%%%

Black body radiation and synchrotron radiation are two extremely important
phenomena in the realm of physics. The first one is under investigation for
more than one hundred years, whereas the latter \cite{B} was first observed
in 1947, two years after the discovery of the synchrotron
motion \cite{mmv}.
They might be considered as completely different simply because the
usual significance of their names is opposite: thermal radiation and
nonthermal (magnetobremsstrahlung) radiation. However the progress in
a number of areas of quantum field theories, first of all as related to
accelerated quantum detectors (Unruh effect), challenged me to undertake
the small steps in the following
toward reviewing several interconnected topics in order to point out the
relationships
between these two radiation patterns. More exactly, one can still have doubts
on the 
empirical equivalence between thermodynamic (Boltzmann)
temperature and
KMS quantum field temperature (merely an analytic continuation property),
though one can look at various
bremsstrahlungs as {\em temperature} radiation in the KMS sense.
%In my opinion this is not only a curious, equivalent viewpoint.
This is an interpretation
that may turn useful in extending by far the present limits of
radiometric
standards to the radiation patterns in an accelerator context and even to the
much wider astrophysical context \cite{pis}. Moreover, even the
black holes can be considered in the radiometric perspective \cite{ro1}.
Besides, one may hope that the synchrotron radiation, as well as other
bremsstrahlung patterns, can be
obtained from the `thermal' spectrum by some mathematical procedure, e.g.,
by means of quantum-deformations. This is only one {\it thesis} that will not
be pursued in the following, instead various other issues are to be encountered
herein. For example, in the literature there have been discussed
connections between the `circular' Unruh effect and synchrotron
radiation. Indeed, from the standpoint of quantum field theory, the Unruh
effect, and
also the Hawking effect, can be considered as manifestations of Bogolubov
transformations modifying the structure of zero-point noise. One can
call {\em Bogolubov vacuum noises} all these noises. Hawking and Rindler
noises are Bogolubov noises possessing pure `thermal' (temperature) character,
but this is not a general rule. This may suggest to interpret the synchrotron
radiation as a Bogolubov noise, or at least a part of it, that one
corresponding to stationary trajectories in the sense of Letaw (see below).
A few steps in this direction have been made
in the literature. Hacyan and Sarmiento \cite{hs} calculated the vacuum
stress-energy tensor of the electromagnetic field in a rotating frame and
found a nonzero energy flux in the direction of motion of the observer
(the electron) in such a frame. This prompted Mane \cite{ma} to speculate
that the Hacyan-Sarmiento flux is the synchrotron radiation, but clearly
more details have to be worked out.

In principle, there is essentially nothing new in these notes, which have been
written in order to provide a sort of background material for further studies.
%a construction of mine that I did in order to put some order and to filter
%through writing the flux of information.
They are organized in the nine sections mentioned in their contents.
 
\newpage
%%%%%%%%%%%%%%%%%%%%%%%%%%%%%%%%%%%%%%%%%%%%%%%%%%%%%%%%%%%%%%%%%%%%%%%%%%
%\section{Synchrotron radiation}  %%%% 2
%%%%%%%%%%%%%%%%%%%%%%%%%%%%%%%%%%%%%%%%%%%%%%%%%%%%%%%%%%%%%%%%%%%%%%%%%%%
\section{Magnetobremsstrahlung radiation patterns}

Magnetobremsstrahlung is well known in the synchrotron regime,
in which the angle $\Delta \alpha$ by which the velocity of a relativistic
point electron rotates in the magnetic field is much greater than the angle
$\Delta \theta=1/\gamma$ in which the main part of the radiation pattern is
contained. The opposite regime, on the other hand, is called undulator
radiation. Vice versa, one can consider the synchrotron radiation as a
special case of the undulator radiation \cite{1}. In more definite terms,
one can use the parameter $K\approx \frac{\lambda_u e B_1}{\pi m_0 \beta c}$
characterizing the insertion devices at storage rings, where
$B_1=\frac{B_0}{\cosh (\pi g/\lambda _u)}$, with $B_0$ the field at the pole
surface, $g$ the gap of the insertion device, and $\lambda _u$ the undulator
wavelength. The exact formula for the $K$-parameter contains a numerical
factor expressing the geometrical configuration of the device, i.e., the
type of the beam trajectory in the device. In the undulator magnet one can
define a beam orbit as the trajectory of the centre of motion. The particle
trajectory usually makes an angle with respect to the orbit and it is this
angle that is providing the angular opening of the radiation. It can be
written in the form $\Delta \theta = \frac{K}{\gamma}$ and consequently
one can make the following $K$-classification of the radiation patterns:

$K<1\;\;\;\;$ undulator radiation

$K=1\;\;\;\;$ synchrotron radiation

$K>1\;\;\;\;$ wiggler radiation

The spectral intensity of the magnetobremsstrahlung in the synchrotron
regime reads
%%%%%%%%%%%%%%%%%%%%%%%%%%%%%%%%%%%%%%
$$ W_{\omega}=\frac{9\sqrt{3}}{8\pi}\frac{1}{\omega}
F\Big(\frac{\omega}{\omega _c}\Big)
\eqno(1)
$$
%%%%%%%%%%%%%%%%%%%%%%%%%%%
where $\omega _c=\frac{3}{2}\frac{eH}{mc} \gamma ^2$ and
$F(\xi) =\xi \int _{\xi} ^{\infty}K_{5/3}(z)dz$, where $K$ is the
MacDonald function of the quoted fractional order. The asymptotic limits of
the shape function are as follows
%%%%%%%%%%%%%%%%%%%%%%
$$ F(\xi \ll 1)\approx \frac{2^{5/3}\pi}{\sqrt{3}\Gamma (1/3)}\xi ^{1/3}
\eqno(2)
$$
%%%%%%%%%%%%%%%%%%%%%%
and
%%%%%%%%%%%%%%%%%%%%
$$F(\xi \gg 1)\approx \sqrt{\pi/2}\xi ^{1/2}e^{-\xi}
\eqno(3)
$$
%%%%%%%%%%%%%%%%%%%%%%
The maximum amount of radiation is to be found at the frequency
$\omega _{m}\approx 0.29\omega _c$.

It is well known that much of the analytical structure of the synchrotron
radiation is based on the classical Schott formula of 1907 giving the
spectral and angular features of the differential power emitted by a point
electron moving on a circular trajectory in a constant magnetic field. This
is an exact result based on the classical hypothesis that the electron is
radiating electromagnetic waves in a continuous manner. The Schott formula
reads \cite{s07}
%%%%%%%%%%%%%%%%%%%
$$
W(\nu ,\theta)=\frac{e^2\beta ^2 c}{2\pi R^2}\nu ^2\Big[{\rm ctg}^2\theta
J_{\nu} ^{2}(\nu \beta \sin\theta) +
\beta ^{2}J_{\nu} ^{'2} (\nu \beta \sin\theta)\Big]
\eqno(4)
$$
%%%%%%%%%%%%%%%%%%%%
where $\nu$ is the harmonic number and $\theta$ is the emission angle.
Integrating over angles, and taking into account the asymptotic expressions
for the Bessel function and its derivative, one will find out the
nonrelativistic radiation
%%%%%%%%%%%%%%%%%%
$$
W_{NR}(\nu)=2\frac{e^2 c}{R^2}\beta^{2(\nu +1)}(\nu +1)
\frac{\nu ^{2\nu +1}}{(2\nu +1)!}
\eqno(5)
$$
%%%%%%%%%%%%%%%%%%%%%%%
One can readily see the dipole character of this nonrelativistic radiation
pattern because the maximum is clearly located at the (first) fundamental
harmonic ($\nu$= 1).

On the other hand, at relativistic energies one will find out large shifts
of the maximum from the dipole fundamental harmonic value to the synchrotron
value $\nu \approx \gamma ^3$. Mathematically, this is a consequence of the
asymptotic behavior of the Bessel functions. The characteristic emitting
element of a circular trajectory as seen from the laboratory frame is
$l_{sync}=R\gamma ^{-1}$.

%%%%%%%%%%%%%%%%%%%%%%%%%%%%%%%%%%%%%%%%%%%%%%%%%%%%%%%%%%%%%%%%%%%%%%%%%%%%
\section{Quantum criteria in synchrotron radiation}   %%%% 3
%%%%%%%%%%%%%%%%%%%%%%%%%%%%%%%%%%%%%%%%%%%%%%%%%%%%%%%%%%%%%%%%%%%%%%%%%%%%
%{\bf 3. Quantum criteria in synchrotron radiation}

This section is devoted to an old and quite well-settled issue that I included
for the self-consistency of the paper.
The problem is what are the scales at which quantum effects start showing
up in synchrotron radiation. Two energy scales were thought and elaborated
upon at the end of the forties.

{\bf 3.1 The one half criterion}

The so-called $E_{1/2}$ criterion means
the photon energy corresponding to the maximum of the synchrotron
spectrum be equal to the electron energy
$\epsilon=\frac{\hbar c}{R}\gamma ^3 \geq E$. The equality takes place at
$E_{1/2}=mc^2(\frac{mcR}{\hbar})^{1/2}$ or in terms of the $\gamma$ factor
at $\gamma _{1/2}=(R/\lambda _c)^{1/2}$.

This criterion corresponds to a scaling covariant parameter
$\Upsilon=(H/H_{cr})\gamma=(E/E_{1/2})^2$ where
$H_{cr}=m^2c^3/e\hbar=4.41\times 10^{13} G$ is the Schwinger critical
value of the magnetic field. The quantum synchrotron regime implied by
the one-half criterion is 10 TeV and is not available at the present
accelerators.

%\underline{
{\bf 3.2 The one fifth criterion}
%}

This is a quantum recoil criterion.
The philosophy for this weaker criterion is the following one. The quantum
synchrotron emissions are quite efficient in turning a classical trajectory
into a noisy one. The number of hard synchrotron photons emitted in one
period is $N_1\approx 4\alpha \gamma$. The pathlength of the electrons
without emitting hard photons is $L_0\approx 35(10^4/H)$, where $H$ is in
gauss. The effect of the quantum recoil of the hard photons on the electron
trajectory is given by a more accesible energy scale
$E_{1/5}=mc^2(\frac{mcR}{\hbar})^{1/5}$ or
$\gamma _{1/5}=(R/\lambda _c)^{1/5}$. It may be termed a quasi-quantum
regime in the sense that the global description of the radiation remains
classical to a good approximation, while the trajectory is affected by
the synchrotron quantum fluctuations. In other words, the quantum recoils
are not manifest in the radiated power; firstly, they occur as a trajectory
spreading effect that can be expressed through the following
{\it synchrotron variance}
%%%%%%%%%%%%%%%%%%%%
$$
<\Delta r^2>=\frac{55}{48\sqrt{3}}\frac{e^2}{mc}
\frac{\lambda _c}{R}\gamma ^5t
\eqno(6)
$$
%%%%%%%%%%%%%%%%%%%%%%
One immediately can notice the analogy with the Brownian motion. The quantum
diffusion coefficient for the quantum trajectory effect in motion with
synchrotron radiation is the following
%%%%%%%%%%%%%%%%%%%%%%%%%%%%
$$
D_{s}=\frac{55}{24\sqrt{3}}\frac{e^2}{mc}\frac{\lambda _c}{R} \gamma ^5
\eqno(7)
$$
%%%%%%%%%%%%%%%%%%%%%%%%%%%%
The synchrotron quantum diffusion is a radial diffusion with respect to the
mean classical
circular trajectory. However, we have to pay attention to the fact that in
the case of synchrotron radiation from cyclic accelerators there is another
classical parameter by which one takes into account the effect of the magnetic
field inhomogeneities \cite{js}. It is the following one
$g_{h}=\sqrt{q}\frac{A}{R_{eq}}\gamma ^{-1}$, where $A$ is the amplitude of
the vertical oscillations, $q$ is the inhomogeneity parameter of the magnetic
field, and $R_{eq}$ is the radius of the equilibrium circular orbit. The small
oscillations of the electron around $R_{eq}$ lead to a sharp dependence of
the spectral angular and polarization characteristics of the radiation on
the amplitude of the oscillations.

%\underline{
{\bf 3.3 Quantum regimes in the beamstrahlung case}
%}

The treatment of various quantum regimes becomes more intricate in the case
of beamstrahlung, a well-known by now synchrotron-type radiation at linear
colliders. The beamstrahlung occurs whenever two bunches pass through each
other. The particles in one bunch are deflected by the fields in the
opposing bunch, and as a result a large number of hard photons are produced.
Jacob and Wu \cite{jw} have discussed the length scales
which compete to settle the quantum (radiative) processes of the
e$^{\pm}$-bunch interactions. The first is the {\em correlation length}, i.e.,
the distance travelled to get an angular deflection of $\gamma ^{-1}$.
The second one is the {\em bunch length}, which in the laboratory system is
the bunch longitudinal size $\sigma _{x}$. Finally, the third one is the
so-called {\em virtual electron} length. This is the distance over which the
electron is
radiating coherently before entering the bunch. It is a coherent radiation
which is due to the quantum mechanical transfer of momentum from the bunch,
having a spatial extent increasing linearly with $\gamma$. At the same time,
the discussion of localization is equivalent with a better understanding
of the delocalization of photons in the case of beamstrahlung.

%\newpage

%\underline{
{\bf 3.4 Lorentz-Dirac and quantum Langevin equations}
%}

Since the quantum effects in the radiated power are not important at the
common accelerator energies an effective equation of motion often used for
a single electron is the classical Abraham-Lorentz-Dirac (ADL) equation
%%%%%%%%%%%%%%%
$$
m\ddot x ^{\mu}=\frac{e}{c}x_{\nu} {\cal F ^{\mu \nu}}+\frac{2}{3}
\frac{e^2}{c^3}\left(\stackrel{...}{x} ^{\mu}-
\frac{1}{c^2}\dot x^{\mu}\ddot x^{\nu}\ddot x_{\nu}\right)
\eqno(8)
$$
%%%%%%%%%%%%%%%%%%%
The derivatives are taken with respect to the proper time. The last term is
the friction term due to the radiation reaction, and it is precisely the third
derivative that turns this equation into a rather special case within
the whole range of equations in mathematical physics. It is one of the most
controversial equations of physics, showing up famous shortcomings, like
runaway solutions and preacceleration, that have been a continuous sourse
of discussions in the literature. The (A)LD equation can be considered as
the exact equation of motion for a point charge in classical electrodynamics.
At the classical field level the (A)LD equation is a reasonable way of
taking into account the dynamical feedback of the radiation on the particle
dynamics, i.e., the radiation reaction, whenever we apply an external field
to the point charge.

In the quantum approach the particle dynamics has to be considered as
distributed over the energy states of the electron in the external magnetic
field and therefore the damping action of the radiation reaction is
automatically included in the dynamics. However the spreading of the electron's
motion over the energy states is strongly dependent on the electron energy,
just because the emission of a hard photon is not equally probable to that
of a soft one. Considering the interaction Hamiltonian as a small perturbation
to the Dirac equation in a stationary magnetic field, one will find for the
electron the well-known Landau levels
$E_n=mc^2[1+(P_3/mc)^2 +2H/H_{cr}]^{1/2}$. These levels, depending only on
the total quantum number $n$ are degenerate. The transition probability
between two unperturbed energy levels reads as follows
%%%%%%%%%%%%%%%%%%%%%
$$
\lambda _{ij}=\frac{\alpha \omega _{H}}{\sqrt{3}\pi}
\frac{(mc^2)^3}{E_i^{2}E_j}f(y)
\eqno(9)
$$
%%%%%%%%%%%%%%%%%%%%%%%%%%
where a summation over the degenerate states should be supposed and $f(y)$
is the following integral
%%%%%%%%%%%%%%%%%%%
$$
f(y)=\int _{y}^{\infty}K_{5/3}(x)dx + (3/2 \Upsilon y)^2
(1+3/2\Upsilon y)^{-1}K_{2/3}(y)
\eqno(10)
$$
%%%%%%%%%%%%%%%%%%%%%%%%%%%%
where the variable $y=(E_i-E_j)/R_{qi}E_j$ and $R_{qi}=3/2\gamma _i H/H_{cr}=
\Upsilon _i  $. The transition rates $\lambda _{ij}$ vary as $y^{-2/3}$ for
$y\ll 1$ and drops off exponentially for $y\gg 1$. In this framework the
radiation spectrum of an electron at energy $E_i$ is given by
$\sum _j \hbar \omega _{ij}\lambda _{ij}=I_{i}(\omega)d\omega$, where
$I_{i}$ is expressed by the rather long formula
$I_i=(3\sqrt{3}/4\pi)c\gamma _i ^2 H^2 (e^2/mc^2)^2(z^2/(1+3/2\Upsilon _i z)^3)
f(z)/\omega(1-\hbar\omega /\gamma _i mc^2)$, and $z=\hbar \omega/\frac{3}{2}
\Upsilon _i (E_i-\hbar \omega)$.

In the relativistic case the third derivative term in the (A)LD equation
can be neglected. Thus, the equation takes the effective form
%%%%%%%%%%%%%%%%%%%%%%%%
$$
m\ddot x^{\mu}=-\frac{e}{c}\dot x_{\nu}{\cal F}^{\mu \nu}-(W/c^{2})
\dot X^{\mu}
\eqno(11)
$$
%%%%%%%%%%%%%%%%%%%%%%%%%%
where $W=\frac{2}{3}\frac{e^{2}c}{R^{2}}\gamma ^{4}$.
In a three-dimensional writing the equation turns into
%%%%%%%%%%%%%%%%%%%%%%%%
$$
\frac{d}{dt}\left(\frac{E}{c^2}\vec v\right)=-\frac{e}{c}\vec v\times \vec H
-e{\cal E}-(W/c^2)\vec v
\eqno(12.a)
$$
%%%%%%%%%%%%%%%
$$
\frac{dE}{dt}=-e(\vec v {\cal E})-W
\eqno(12.b)
$$
%%%%%%%%%%%%%%%
In this more familiar form one can understand straightforwardly the law of
energy conservation: the high frequency electric field compansates the
radiative losses. Suppose one considers the electron motion in a focusing
magnetic field of the common accelerator type
%%%%%%%%%%%%%%%%%%%
$$
\vec H=(-\frac{xzq}{r^{2+q}}b,\; -\frac{yzq}{r^{2+q}}b,\;\frac{1}{r^q}b)
\eqno(13)
$$
%%%%%%%%%%%%%%%%%%%
where $b=H_{0}R^{q}$ and $0<q<1$ is the focusing magnetic coefficient. Plugging
Eq.~(13) in Eq.~(12.a) and using the linear approximation one will
get the usual harmonic betatron and phase oscillations. These
equations display an adiabatic damping corresponding to the growth of the
particle energy in the accelerator and a damping due to radiation reaction.

One can think of the microscopic bases of the electromagnetic radiation
patterns as derived from the Heisenberg picture of Langevin equations.
A general form of the equation of motion for a time dependent operator
$\Omega$ can be written down as follows
%%%%%%%%%%%%%%%%%%%%
$$
\frac{\partial \Omega}{\partial t}=\frac{i}{\hbar}[H,\Omega] +\;damping\;
+\; fluctuations
\eqno(14)
$$
%%%%%%%%%%%%%%%%%%%%%%
According to Haken \cite{hak} such an equation can be put upon rearrangements
in the form
%%%%%%%%%%%%%%%%%%%%
$$
\frac{\partial \Omega}{\partial t}=N(\Omega , \alpha) +\; F(t)
\eqno(15)
$$
%%%%%%%%%%%%%%%%%%%%
where $N$ is a nonlinear function of the operator $\Omega$, depending also
on the set of control parameters $\alpha$ by which one usually describes the
amount of power pumped into a system. $F$ represents the fluctuating forces
which one will like to introduce to include stochasticity. Langevin equations
are the natural way of including quantum effects on the electron trajectory.

%to be completed

%%%%%%%%%%%%%%%%%%%%%%%%%%%%%%%%%%%%%%%%%%%%%%%%%%%%%%%%%%%%%%%%%%%%%%%%%%%%%%
\section{Trajectories (worldlines) and radiation patterns from
quantum detectors}                                             %%%%%% 4
%%%%%%%%%%%%%%%%%%%%%%%%%%%%%%%%%%%%%%%%%%%%%%%%%%%%%%%%%%%%%%%%%%%%%%%%%%%%%%
%{\bf 4.
%Trajectories (worldlines) and radiation patterns from quantum detectors}

The one-fifth quantum criterion shows us that at the current accelerator
energies the quantum effects are of recoil type on the classical trajectory.
Suppose therefore that, in relativistic terms, we stick to the concept of
trajectory/worldline in a given field vacuum, and we would like to investigate
the radiation pattern generated during the movement of a quantum
particle/detector. Since we will speak of the excitation of the quantum
detector, we'll come across the concept of the quantum autocorrelation
function, which is the Wightman function for the vacuum field through which
the particle is propagating. It is well-known that autocovarriance functions
(ACVF)
completely characterize mean square continuous second-order stationary
stochastic processes. Only for stationary processes it is possible to write
the
spectral density directly in terms of ACVF by a formula of the usual type
%%%%%%%%%%%%%%%%%%%%%%%%%%
$$
g(\omega)=2 \int _{0}^{\infty} \cos (\omega \tau) \gamma (\tau)d\tau
\eqno(16)
$$
%%%%%%%%%%%%%%%%%%%%%%%%%%%%%%
where $\gamma (\tau)$ is the Wightman ACVF and $\tau$ is the proper time.
The spectral density is in physical terms the rate of excitation of the
quantum detector induced by the vacuum fluctuations through which is
propagating. When the spectral density is multiplied by the density of
states one will get a power spectrum that one may call (following, e.g.,
Takagi) the vacuum power spectrum. 
A classification of stationary worldlines, which are of helical-type,
has been performed by Letaw in
terms of geometric invariants of the trajectories by means of a
generalization of the Frenet equations to Minkowski space \cite{let}.
However quantum vacua are not
stationary stochastic processes for all types of classical trajectories, in
which case joint time and frequency information is required for tomographycal
processing \cite{tom}.
On the other hand,
it is known from mathematics that certain stationary increment random fields
still posses covariant spectra. The stationary increment processes have
the same sort of asymptotic statistical properties. In an even more
complicated case, in which chaotic trajectories are superposed on more
regular one, like in synchrotron radiation, one has to think about how to
characterize the chaotic trajectories from the noise standpoint.\\
The criterion for the Letaw classification of worldlines is the
time-independence of the excitation spectra of the quantum detectors moving
in the vacuum field. The classification scheme is according to the
curvature invariants of the curve (worldline) which are the proper
acceleration and the angular velocity. The stationary worldlines are
constructed by making use of a generalization of the Frenet equations to
Minkowski space. The basic idea of the classification resides in the fact
that Wightman ACF's for a scalar field are essentially the inverse of the
geodetic interval, and as such, in order to define a power spectrum one
needs the geodetic interval to depend only on the corresponding proper time
interval. Letaw has obtained six types of stationary worldlines as solutions
of the generalized Frenet equations under the condition of constant
curvature invariants. Moreover, he proved the equivalence of the timelike
Killing vectorfield orbits and the stationary worldlines. At this point
I would like to recall the difference between a streamline and a trajectory
in the fluid mechanics, following the textbook of Abraham, Marsden, and Ratiu
[AMR] \cite{amr}. A streamline at a fixed time $t$ is an integral curve of
the velocity field $u(x,t)$. That means that if a streamline $x(t)$ is
parametrized by a parameter $s$ at the instant $t$, then $x(s)$ satisfies
the relation $\frac{dx}{ds}=u(x(s),t)$ for a fixed $t$. On the other hand,
a {\em trajectory} is the curve traced out by the particle/detector as time
is evolving, i.e., it is a solution of the differential equation
$\frac{dx}{dt}=u(x(t),t)$ with given initial conditions. The two concepts
of streamline and trajectory coincide only when the velocity field is time
independent, that is for stationary (hydrodynamical) flows. This means
the geometry of the fluid flow is not changing in time. Similarly, the
geometric properties of Letaw worldlines are independent of the proper
time. Letaw constructed orthonormal tetrads at every point on the worldline,
which are formed of the derivatives of the worldline with respect to the
proper time. A Schmidt orthogonalization procedure is applied to these
first-order derivatives, which are turned into the basis for a vector space
at each point on the worldline. The generalized Frenet relations are nothing
else but the expansion of the derivatives of the basis vectors in terms of
the basis $\frac{dV_a^{\mu}}{d\tau}=K_{a}^{b}V_b^{\mu} $, where $\tau$ is
the proper time and the Latin index is the tetrad index. The $K$ matrix is
an antisymmetric matrix with the entries given by the curvature, torsion,
and hypertorsion of the worldline, $\kappa _i$, $\tau _i$, and $\nu _i$,
respectively, where the subscript denotes the point on the worldline.

%%%%%%%%%%%%%%%%%%%%%%%%%%%%%%%%%%%%%%%%%%%%%%%%%%%%%%%%%%%%%%%%%%%%%%%%%%%%
\section{QED bremsstrahlung and vacuum thermal noises}   %%%%%%%%  5
%%%%%%%%%%%%%%%%%%%%%%%%%%%%%%%%%%%%%%%%%%%%%%%%%%%%%%%%%%%%%%%%%%%%%%%%%%%%%
%{\bf 5. QED bremsstrahlung and vacuum thermal noises}

A number of authors have discussed in the past, with various degrees of
detail, the connection between the trajectory (bremsstrahlung) noises and the
vacuum heat-bath interpretation. For a review, see Takagi \cite{tak}.

The first to focus more on the connection between bremsstrahlung and vacuum
baths was Kolbenstvedt \cite{kol}. He considered the case of scalar
bremsstrahlung for the UDW monopole quantum detector. Higuchi, Matsas, and
Sudarsky [HMS] \cite{hms} dealt with this problem too. They showed that
the usual
QED bremsstrahlung from a point charge moving with constant proper
acceleration can be reproduced in the coaccelerated frame as the combined
rate of emission (absorption) of zero-energy Rindler photons into (from)
the Unruh thermal bath. Their model was a simple oscillating dipole
arangement of length $L$ for which they calculated the dipole radiation in
the vacuum bath. In the end they proceeded with two limits as follows. One is
the $L\rightarrow\infty$ limit in order to remain with the radiation of one
charge alone (and as they proved the second charge and the current flow
between the two charges do not contribute to the final result), and the
other limit is the null-energy limit in order to stay in the rest frame.
Recall the line element of the Rindler wedge
%%%%%%%%%%%%%%%%%%%%%%%%%
$$
ds^2=e^{2a\xi}(d\tau ^2 - d\xi ^2 )-dx^2 -dy^2
\eqno(17)
$$
%%%%%%%%%%%%%%%%%%%%%%%%%%%%%
The Rindler coordinates are related to the Minkowski coordinates in the
following well-known manner
%%%%%%%%%%%%%%%%%%%%%%
$$
t=\frac{e^{a\xi}}{a}\sinh(a\tau)
\eqno(18.a)
$$
%%%%%%%%%%%%%%%%%%%%%%%%%%%%
$$
z=\frac{e^{a\xi}}{a}\cosh(a\tau)
\eqno(18.b)
$$
%%%%%%%%%%%%%%%%%%%%%%%%%%
There are three Killing fields helping to look for solutions of a Laplacian
equation in the Lorentz gauge. Up to a scaling the Rindler problem is
identical with the `hyperbolic motion' of a single charge in a constant
electric field, which is the first exact solution in relativistic dynamics
provided by Max Born \cite{bor} in 1909, and shown later by Schott \cite{stt}
to be the only solution of the Lorentz-Dirac equation with zero
radiation reaction. The physical modes of the Rindler problem contain
the following MacDonald function of imaginary order
%%%%%%%%%%%%%%%%%%%
$$
\phi=K_{i\omega /a}({\bf{k}}_{\perp} e^{a\xi}/a)
\eqno(19)
$$
%%%%%%%%%%%%%%%%%%%%%
HMS evaluated the amplitude for the absorption of a Rindler photon by the
accelerated charge by coupling the Rindler vacuum to the
excitable physical modes through an interaction Lagrangian. The probability
of absorption per unit time for fixed transverse momentum $k_{\perp}$ is
%%%%%%%%%%%%%%%%%%%%%%%%
$$
dW^{abs}(\omega , k_{\perp})=|{\cal A}^{abs}|^{2} d\omega /T=
\Big[{\frac{q^2 E}{4\pi ^2 a^2}|K^{'}_{iE/a}({\bf k}_{\perp}/a)|^{2} +
{\cal O} (E^{3})}\Big]\delta (E-\omega)d\omega
\eqno(20)
$$
%%%%%%%%%%%%%%%%%%%%%%%%%%
Taking now into account the Rindler bath, the total absorption rate of Rindler
`photons' at fixed transverse momentum reads
%%%%%%%%%%%%%%%%%%%%%%%%%
$$
P^{abs}({\bf k}_{\perp})=\int _{0}^{\infty} dW^{abs}
(\omega, {\bf k}_{\perp})\; \frac{1}{e^{2\pi \omega /a}-1}
\eqno(21)
$$
%%%%%%%%%%%%%%%%%%%%%%%%%
Performing the integral and taking the null-energy limit, the following
result can be obtained \cite{hms}
%%%%%%%%%%%%%%%%%%%%%%%%%%
$$
P^{abs}_{k_{\perp}}=\frac{q^2}{8\pi ^3 a}|K_{1}({\bf k}_{\perp}/a)|^{2}
\eqno(22)
$$
%%%%%%%%%%%%%%%%%%%%%%%%%%
On the other hand the total emission rate at fixed transverse momentum can be
written down as
%%%%%%%%%%%%%%%%%%%%%%
$$
P^{em}_{k_{\perp}}=\int _{0}^{\infty}dW^{em}(\omega, {\bf k}_{\perp})
\Big[\frac{1}{e^{2\pi \omega /a -1}} +1\Big]
\eqno(23)
$$
%%%%%%%%%%%%%%%%%%%%%%%%%
where the first term is the induced emission while the second is the
spontaneous emission. Since $dW^{em}=dW^{abs}$ by unitarity, one can use
$dW^{abs}$ to integrate in the last formula, and in the null-energy limit
one gets $P^{em}_{k_{\perp}}=P^{abs}_{k_{\perp}}$. The cause of nonvanishing
probabilities in the null-energy limit is the infinite number of zero-energy
Rindler `photons' of the vacuum heat-bath. The total Rindler rate will be
the sum of the two rates, i.e., the double of each of them.
HMS provided also a calculation of the emission rate at fixed transverse
momentum in the inertial frame (a boost invariant), obtaining the same
result as in the Rindler frame.

Before ending this section, let's touch upon the KMS condition which is
essential for vacuum heat-baths. In quantum theory of finite degrees of
freedom (quantum mechanics) the equilibrium states are characterized in the
usual statistical way
$\langle A\rangle=Tr(e^{-\beta H} A)/Tr e^{-\beta H} $
where $H$ is the Hamilton operator and $A$ a Heisenberg observable.
They are equally well described by KMS-type of boundary conditions
$\langle A_{\tau}B\rangle=\langle BA_{\tau +i\beta} \rangle$ where the
subscript
denotes the time translated observable. In quantum field theory only the
KMS condition can be used to characterize {\em thermality} \cite{hhw} since
the statistical definition does not exist. In 1982, Sewell wrote a seminal
paper \cite{sew} in which he explicitly illustrated the manner in which the
PCT symmetry and KMS features of relativistic quantum field theory are
related to Hawking effect and Unruh effect. As a matter of fact, a simple
derivation of the KMS condition is to start with the following definition
$\langle A(t)B(t^{'})\rangle=Tr(A(t)B(t^{'})\rho)$ where $\rho$ is the density
matrix for a state of temperature T in quantum field theory, and to substitute
$A(t-i\beta)$ instead of $A(t)$, the cyclic invariance of the trace, and the
identity $A(t-i\beta)=e^{\beta H}A(t)e^{-\beta H}$.

The following nice argument has been used by Bell, Hughes, and Leinaas
\cite{bhl}. Suppose a thermometer, i.e., a weakly coupled system, is added
to another system (a quantum field) that one would like to probe in a
thermodynamical way. One can show that the transition rates for the direct
and inverse processes in the thermometer system in lowest order perturbation
theory are related as follows $R_{if}=e^{\beta (E_f -E_i)}R_{fi}$. To get this
relation, the KMS condition is essential. In physical terms it means that an
incoherent mixture of energy eigenstates of the thermometer will be in
equilibrium with the system iff the occupation probabilities $\rho _{i}$ and
$\rho _{f}$ of its states are in the ratio $\rho _{f}/\rho_{i}=
e^{-\beta (E_f-E_i)}$. This implies that the thermometer is at the same
temperature as the quantum field. Now, when one is dealing with the analog
weak coupling equilibrium for the case of uniformly linear accelerated
detector, the KMS condition must be written in the form
$\langle A(\xi , \tau-i\beta _{U})B(\xi ^{'}, 0)\rangle=
\langle B(\xi ^{'},0)A(\xi ,\tau)\rangle  $
where the spacetime coordinates of the operators are Rindler ones and the
expectation values are taken over the Minkowski vacuum; $\beta _{U}$ is the
Unruh beta parameter and in the weak coupling limit is common both to the
thermometer and the field. More rigorously, the KMS condition describes
equilibrium dynamical states generated by the automorphisms of a $C^*$
algebra. The celebrated Bisognano-Wichmann theorem states the following:
{\em the boosts
satisfy the KMS condition with respect to the algebraically defined vacuum
as automorphisms of
the von Neumann algebras of the corresponding wedge region in Minkowski
spacetime} \cite{bw}. More details and references can be found in the book of
Haag \cite{haa} and also in the lectures of Petz \cite{petz}, but
the best reference is the very recent Los Alamos archive entry of
Schroer \cite{s97}.
The connection between the mathematical theory of modular covariance
\cite{take,sz}
and the physics of Hawking effect has been first discussed by
Sewell \cite{sew}.

%%%%%%%%%%%%%%%%%%%%%%%%%%%%%%%%%%%%%%%%%%%%%%%%%%%%%%%%%%%%%%%%%%%%%%%%%%%
\section{Radiometric characterization of synchrotron radiation}  %%%%  6
%%%%%%%%%%%%%%%%%%%%%%%%%%%%%%%%%%%%%%%%%%%%%%%%%%%%%%%%%%%%%%%%%%%%%%%%%%%%
%{\bf 6. Radiometric characterization of synchrotron radiation}

The real motivation of these notes is my hope that the progress achieved
in the interpretation of the ``classical"-trajectory propagation of quantum
particles in quantum field vacua can have a definite
impact on such important topics as extending the limits of radiometry by
a more precise characterization of quantum field radiation standards.
I have already expounded a little on this issue elsewhere \cite{ronc}, and
I consider it as the main application of the highly acclaimed theoretical
results of Hawking and Unruh.

Usually, the bunches circulating in a storage ring contain about 10$^{11}$
electrons. Only a small part of them will follow stationary orbits in the sense
of Letaw. For these Letaw electrons an effective KMS temperature may be taken
into account for the beam. Hopefully, for the rest of the electrons,
radiometric considerations will emerge sooner or later depending on the pace of
our understanding of the Hamiltonian chaos, either quantum or classical.
I became also aware of the progress at the Daresbury synchrotron storage
ring (SRS) \cite{srs}. The situation there is as follows. In terms of photons
emitted per second the SRS is more intense than a blackbody source only in the
far-infrared region (below 50 cm$^{-1}$). However, the other advantages, like
its brightness (watts per unit area and solid angle) as well as its precise
time structure are manifest throughout the infrared. The Daresbury group
made a direct radiometric comparison between the SRS and a high-pressure
mercury arc lamp, which is a conventional far-infrared source. From the
brightness measurements one can extract equivalent black-body temperatures
according to the formula $ B(\nu)=2k_{B}\nu ^{2}cT_{bb}$ under the conditions
of the same source aperture and the same area-solid angle product). The
Daresbury group found out that in the longitudinal polarization case the
equivalent $T_{bb}$ is beyond 8000 K over the whole measured frequency range
(10-110 cm$^{-1}$) with a maximum of 20000 K at 20 cm$^{-1}$. For transverse
polarization the equivalent temperatures are one order of magnitude lower
and are continuously decreasing with the frequency from a maximum of about
1600 K at 10 cm$^{-1}$. Moreover, the Daresbury group provided plots of the
power and power ratios of the two radiations and compared the data points
for the power ratios with those calculated by Duncan and Williams \cite{dw}.

%%%%%%%%%%%%%%%%%%%%%%%%%%%%%%%%%%%%%%%%%%%%%%%%%%%%%%%%%%%%%%%%%%%%%%%%%%%%%
\section{The problem of coherence in synchrotron radiation}   %%%%%%%%  7
%%%%%%%%%%%%%%%%%%%%%%%%%%%%%%%%%%%%%%%%%%%%%%%%%%%%%%%%%%%%%%%%%%%%%%%%%%%%%
%{\bf 7. The problem of coherence and synchrotron radiation}

Coherent radiation patterns have many advantages over the incoherent ones.
The most obvious one is an enhancement in intensity by several orders of
magnitude. At the present time coherent effects have been observed in
infrared synchrotron radiation \cite{will,nak}, in undulator radiation
\cite{jeo}, and there exist theoretical discussions on the coherent
bremsstrahlung in colliding beams \cite{gin}.

Curtis Michel \cite{cm} tackled the problem
of a coherent output of synchrotron radiation at electron storage rings.
His arguments go as follows. Take a common ring of a radius of about 12 meters
within which bunches of 10$^{11}$ electrons about 1 cm in length are
circulating at energies in the range of several GeV. The resulting synchrotron
radiation occurs mainly over the range of several keV X-ray region and is
incoherent since the electrons are located randomly at many X-ray wavelengths
from one to another. Take now into account that the synchrotron radiation is
usually pulsed because of the bunch structure of the beam and besides is
forward-angular confined because of the relativistic motion of the electrons.
The remark of Curtis Michel was that the flux density in the synchrotron
spectrum falls rather slowly toward low frequencies (as $\omega ^{1/3}$)
remaining relatively high even at wavelengths as long aa the size of the bunch.
Going from X-ray wavelengths to the bunch size (1 cm) the flux density
reduces by less than one thousand times.
Also, Kaltchev and Perelstein \cite{kp} dealt with the coherent corrections
to the radiation spectrum of a relativistic electron beam propagating along
an external magnetic field.

%%%%%%%%%%%%%%%%%%%%%%%%%%%%%%%%%%%%%%%%%%%%%%%%%%%%%%%%%%%%%%%%%%%%%%%%%%%%%
\section{Bremsstrahlung and continued fractions}        %%%%%%%%%%  8
%%%%%%%%%%%%%%%%%%%%%%%%%%%%%%%%%%%%%%%%%%%%%%%%%%%%%%%%%%%%%%%%%%%%%%%%%%%%%%
%{\bf 8. Bremsstrahlung and continued fractions}

A paper written by Fried and Eberly in 1964 is noteworthy \cite{fe}. The
paper deals with the Thomson scattering of high-intensity, low-frequency,
circularly polarized electromagnetic waves by unbound electrons. They obtained
semiclassical solutions by a graph summation technique in which certain
infinite continued fractions are involved. The scattering amplitude is
written down as a sum over all partial scattering operators, expressed by
means of continued fractions. The quantity of interest is just the infinite
continued fraction obtained in the limit. Finally Fried and Eberly employed
theorems due to Van Vleck to express the physical result in terms of
Bessel functions.

%to be completed
\newpage
%%%%%%%%%%%%%%%%%%%%%%%%%%%%%%%%%%%%%%%%%%%%%%%%%%%%%%%%%%%%%%%%%%%%%%%%%%%%%
\section{Conclusions}                                    %%%%%%%%%  9
%%%%%%%%%%%%%%%%%%%%%%%%%%%%%%%%%%%%%%%%%%%%%%%%%%%%%%%%%%%%%%%%%%%%%%%%%%%%%
%{\bf 9. Conclusions}

I presented an intent toward a matching of ideas vehiculated in quantum
field theories as related to accelerated detectors, and features of radiation
patterns in the realm of accelerator physics. There are many promises if
one is decided to embark upon a project of this kind of which seemingly
I was able to
grasp only a tiny introductory part. This might be not only what one
usually classifies as an {\em interesting} project but also a
quite {\em useful} one.
%In a short lapse of time
%I became aware (merely by random quick browsing)
%of a vast literature probably somewhat surpassing my ability of hinting upon
%the essential points.
My guess is that quantum field radiometry at storage
rings will be much in gain from this mixed approach I was trying to present
and I do hope that
some people will take the enterprise. The difficulty consists in fitting
two quite different languages, the quantum field theoretic and the
plain accelerator one.

Further extensions of the issues focused on in these notes regard
the related topics of stochasticity and chaos at storage rings \cite{stoch}.

\nonumsection{Acknowledgements}
\noindent
%\section*{Acknowledgment}
This work was partially supported by the CONACyT Project 4868-E9406.

%This section should come before the References. Funding
%information may also be included here.

\nonumsection{References}
%\noindent
%References are to be listed in the order cited in the text. Use
%the style shown in the following examples. For journal names,
%use the standard abbreviations. Typeset references in 9 pt Times
%Roman.

%\appendix

%\noindent
%Appendices should be used only when absolutely necessary. They
%should come after the References. If there is more than one
%appendix, number them alphabetically. Number displayed equations
%occurring in the Appendix in this way, e.g.~(\ref{that}), (A.2),
%etc.
%\begin{equation}
%\mu(n, t) = {\sum^\infty_{i=1} 1(d_i < t, N(d_i) = n) \over
%\int^t_{\sigma=0} 1(N(\sigma) = n)d\sigma}\,. \label{that}
%\end{equation}
\end{document}